\documentclass[twocolumn,english,aps,prx,reprint,superscriptaddress,showpacs,longbibliography]{revtex4-1}
\usepackage{amsmath,amsfonts,amsthm,amssymb,mathrsfs,bbm}
\usepackage{graphicx,mathptmx}
\usepackage[colorlinks=true,linkcolor=blue,citecolor=blue]{hyperref}
\usepackage{subfigure}
\usepackage[T1]{fontenc}
\usepackage[latin9]{inputenc}
\usepackage{amstext}
\usepackage{float}
\usepackage{color}
\usepackage{enumitem}   

\makeatother

\begin{document}
\title{{Turbulence-induced rogue waves in Kerr resonators}}


\author{Saliya Coulibaly}
\email{saliya.coulibaly@univ-lille.fr}\affiliation{Universit\'e de Lille, CNRS, UMR 8523 - PhLAM - Physique des Lasers Atomes et Mol\'ecules, F-59000 Lille, France.}
\author{Majid Taki} 
\affiliation{Universit\'e de Lille, CNRS, UMR 8523 - PhLAM - Physique des Lasers Atomes et Mol\'ecules, F-59000 Lille, France.}
\author{Abdelkrim Bendahmane}
\affiliation{Laboratoire Interdisciplinaire Carnot de Bourgogne, UMR6303 CNRS - Universit\'e Bourgogne Franche-Comt\'e, Dijon, France.}
\author{Guy Millot}
\affiliation{Laboratoire Interdisciplinaire Carnot de Bourgogne, UMR6303 CNRS - Universit\'e Bourgogne Franche-Comt\'e, Dijon, France.}
\author{Bertrand Kibler}
\affiliation{Laboratoire Interdisciplinaire Carnot de Bourgogne, UMR6303 CNRS - Universit\'e Bourgogne Franche-Comt\'e, Dijon, France.}
\author{Marcel Gabriel Clerc}
\affiliation{Departamento de F\'{i}sica {and Millennium Institute for Research in Optics}, FCFM, Universidad de Chile, Casilla 487-3, Santiago, Chile.}



\begin{abstract}
Spontaneous emergence of self-organized patterns and their bifurcations towards a regime of complex dynamics 
in non-equilibrium dissipative systems is a paradigm of phase transition. Indeed, the behavior of these patterns in the highly nonlinear regime remains less explored, 
even in recent {high-quality-factor} resonators such as Kerr-nonlinear optical { ones}. { 
Here, we investigate theoretically and experimentally the alteration of the resulting Kerr frequency combs from 
the weakly to the highly nonlinear regime, in the frameworks of spatiotemporal chaos, {and} dissipative phase transitions.
We reveal the existence of a striking and easily accessible scenario of spatiotemporal chaos, free of cavity solitons,   
in a monostable operating regime}, wherein a transition to amplitude turbulence via spatiotemporal intermittency 
is evidenced. Moreover, statistics of the light bursts in the resulting turbulent regime unveils the existence 
of rogue waves as extreme events characterized by long-tail statistics.
\end{abstract}

\date{\today}

\maketitle
\section{Introduction}
The concept of order parameter description has played a key role in understanding dissipative structures and self-organized pattern in nonlinear systems. 
{ Indeed, this description has been revealed to be a powerful theoretical tool for classifying and describing 
dissipative structures in weakly nonlinear regimes of nonequilibrium systems as the ones subject to 
a moderate external driving power.  However, high external driving strengths lead dissipative systems 
{to} strongly nonlinear regimes 
where they exhibit extremely complicated dynamics such as spatiotemporal chaos and turbulence. 
In such a case, the degree of complexity of the highly nonlinear problem {excludes} any attempt 
to finding the appropriate order parameter that remains an elusive task. 
The development of a complete process for the study of the complex dynamics, 
including spatiotemporal chaos and turbulence, occurring in highly nonlinear 
regimes of dissipative systems is one of the most challenging open problems in nonlinear 
science \cite{ecke2015, lemoult2016, cardesa2017}. Many recent advances in {the} 
understanding of complex dynamics have been driven by experimental and theoretical studies in modern optics, in fields as diverse as optical fiber cavities with quadratic or cubic nonlinear materials subject to { externally} injected radiations \cite{zhao2006, Turitsyna2013, wabnitz2014}. 
The field of { optics, therefore,} is ideally suited to the investigation 
of spatiotemporal chaos and turbulence in dissipative systems far from thermodynamic equilibrium (cavity and laser systems).}
{Within {a} few decades, optical Kerr-nonlinear resonators have emerged as the {†paradigmatic} 
setup for the study of externally driven nonlinear systems {\cite{Ikeda1979,Ikeda1980,Ikeda1982}}. With length scales ranging from the meter to micrometer~{\cite{Shaw1982,DelHaye2007}}, their applications range from high-demand telecommunications {\cite{Leo2010}} to precision spectroscopy and LIDAR (Light Detection And Ranging) systems {\cite{Savchenkov2016}} based on the coherent optical frequency combs that can be delivered. 
Kerr {resonators} are also known for the property to continuously switch between a monostable (single-valued transmission curve) and bistable (S-shape transmission curve)  regimes. Operating out of equilibrium,  
Kerr resonators can {exhibit nontrivial} outputs such as cavity solitons (localized coherent solution) {\cite{Grelu2015,parra2014,parra2018}} and the modulation instability {\cite{Turing1952, Feir1967}} (process by which a homogeneous state breaks up into a periodic state). Unlike cavity solitons requiring the system to be  bistable,  the modulation instability (MI) was reported both in bistable and monostable regimes. There is currently a renewed interest for the MI in nearly-conservative physical systems after being  tightly linked to the recent research activity and new developments in rogue wave phenomena {\cite{Akhmediev2009}}. Optical analogs of the hydrodynamic rogue waves are rare and short intense light pulses,  characterized by long-tail statistics in the probability distribution of the intensity profile. They have been observed in numerous optical systems and may result from  the emergence {of} coherent states or a collision of them. 
The Peregrine soliton, and Akhmediev  breathers in conservative systems described 
by the nonlinear Schr\"odinger equation are the renowned examples {\cite{Kibler2010}}. Out of equilibrium systems,  to which the ring resonators 
{belong},  have also been reported to exhibit  rogue waves as extreme events.  In this context, they result from deterministic processes leading to complex dynamical evolutions such as temporal chaos {\cite{Lecaplain2012,Pisarchik2011,Bonatto2011,Montina2009}}, spatiotemporal chaos and intermittency {\cite{Selmi2016,Coulibaly2017}. But from the general point of view, the identification of the necessary ingredients for the emergence of rogue waves and extreme 
events in dissipative systems remains {a challenging open problem}, 
including predictability \cite{Steinmeyer2015}, mechanisms of formation, and high sensitivity to noise sources \cite{Mussot2009} of these giant waves.}

{Returning} to Kerr resonators, despite the tremendous interest on their dynamics, 
to the best of our knowledge, the only reference to rogue waves generation is a theoretical observation 
in the bistable regime with anomalous dispersion \cite{Coillet2014}. { Moreover}, more generally, it is MI of the monostable regime which {has} 
received much less attention. However, the MI has been demonstrated to seed an extremely interesting complex dynamics \cite{daviaud1990,Ciliberto1988,gomila2003} when the pertinent control parameter is brought far from the instability onset. Examples include phase instability, spatiotemporal intermittency, and turbulence.

In this work, we investigate experimentally and theoretically the transition from periodic patterns with a triangular comb spectrum induced by 
MI in ring cavities towards more complex dynamics using a combination of three quantities:
\begin{enumerate}[label=({\em \roman*})]
\item {\em The Lyapunov  dimension density}-The proof of the matters of the spatiotemporal chaos in an 
extended  { system} is the existence of a continuous set of positive Lyapunov exponents. {Besides,} 
this set may increase with the number of degree of freedom \cite{Ruelle1982}. Therefore, any quantity 
directly defined from a set of Lyapunov exponents yield to an extensive quantity. For example, 
all definition of the dimension of the attractor (the effective degree of freedom of the system) produces 
a quantity proportional to the volume of the system. The inverse of this proportionality coefficient is an 
intensive quantity with the same dimension of the volume of the system. 
Here, we compute the  Lyapunov  dimension density from the Kaplan-Yorke {conjecture} \cite{Ruelle1982}, 
which  { allows us to determine} the characteristic size of the independent subsystems generated by the spatiotemporal chaos.
\item {\em The {two-point} correlation length}--A consequence of the MI is a diverging correlation length 
(i.e. a correlation that spawn {overall} a finite size system). Hence, any destabilization process of the 
MI may cause a meaningful reduction of the correlation lengths in the system \cite{Cross1993,Egolf1994,Bohr1994,Egolf1995,Ohern1996}. 
In that case, short range correlation length given by the exponential decay rate of the correlation 
function is of particular interest to detect the phase transition-like processes when varying the control parameter of the system.
\item {\em Intrinsic laminar length}--Finite correlation range can be the result {of} a mixed state where spatiotemporal 
coherent (laminar) {subsystems} coexist with incoherent (or chaotic) ones. 
In this case, the probability distribution function of the laminar subsystem gaps can provide a significant signature about the nature of the complex dynamics. Namely, a power law distribution entails the spatiotemporal intermittency and an exponential distribution is a characteristic of the fully developed turbulence. However, it is more likely to find a range of the probability distribution of laminar regions that fits a decaying exponential law \cite{Ciliberto1988,daviaud1990,lemoult2016}. 
We will {refer} to this quantity as the intrinsic laminar length.
\end{enumerate}
Note that, although the use of these quantities is widespread in hydrodynamics \cite{Ciliberto1988,daviaud1990,lemoult2016}, 
only very few examples of their use are known in optics despite the many analogies between these areas.  {Besides}, 
if the correlation length has been already associated with the Lyapunov dimension correlation or to the intrinsic exponential 
decay, the three quantities have not been used together yet. In the following, we numerically demonstrate that a monostable 
Kerr resonator can exhibit spatiotemporal chaotic regime. We have { verify} experimentally and numerically 
that the transition to this a chaotic behavior {coincides} { with} a diverging two-point correlation length in the system 
when varying the external driving strength. The correlation length allows us to identify a second { transition} in the dynamics. 
The numerical computation of the intrinsic laminar length helps to {reveal} this change as the transition from 
spatiotemporal intermittency to fully developed or amplitude turbulence. At this { transition,} 
we have  identified the  { appearance of} extreme events in the intensity profile. 
Finally, the temporal profile of the rogue waves is obtained through the analysis of the correlation function.
{Theoretical results and experimental observations are in  { a} good agreement.}
}
\begin{figure}[b]
\centering 
\includegraphics[width=.5\textwidth]{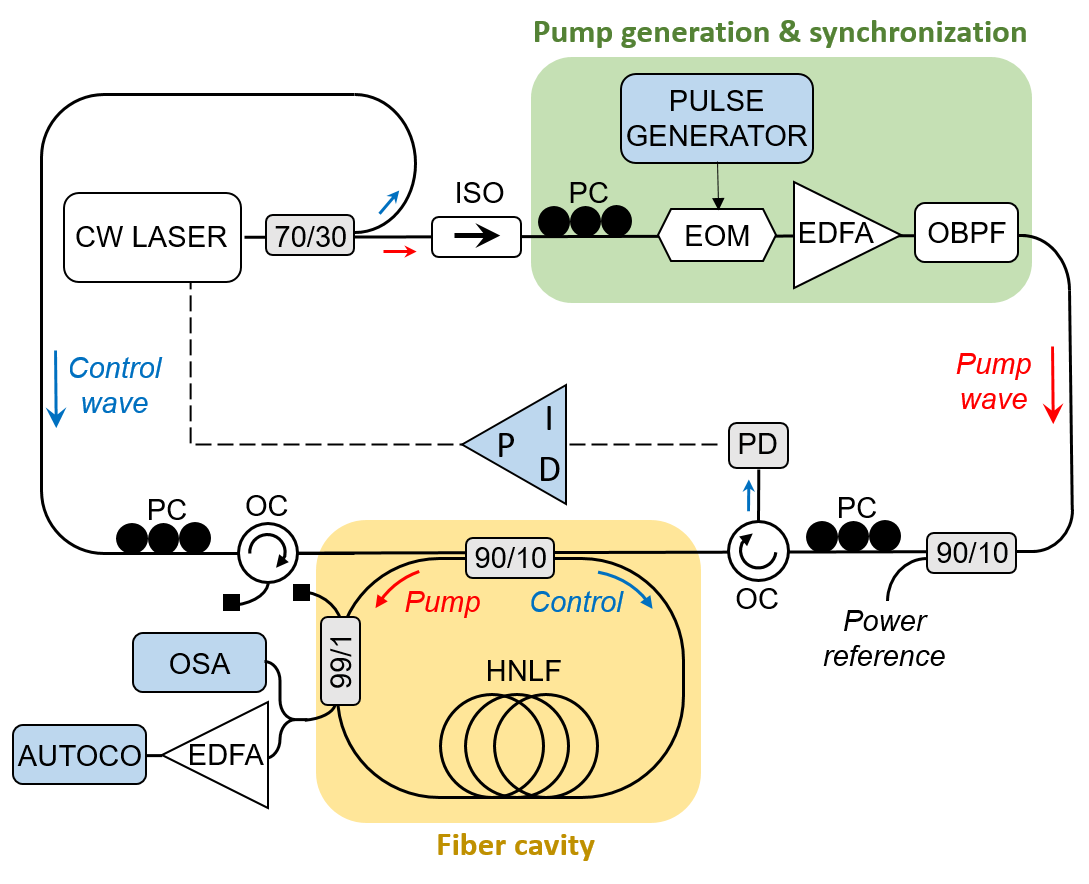}
\caption{Experimental set-up. ISO: { optical} isolator; PC: polarization controller; 
EOM: electro-optic (intensity) modulator; EDFA: Erbium-doped fiber amplifier; OBPF: bandpass optical filter; OC: optical circulator; PD: photodiode; HNLF: highly nonlinear optical fiber; AUTOCO: autocorrelator; OSA: optical spectrum analyzer.} 
\label{fig:ExperimentalSteup} 
\end{figure}

{ The paper is organized as follows. In section II the experimental setup, based on a coherently driven passive optical fibre ring cavity, is illustrated and  the main ingredients for achieving the experiments are described. We also present both the governing equations of the experimental optical device and the reduced Lugiato-Lefever model. In section III spatiotemporal chaos resulting from modulation instability is analytically and numerically studied and the optimal range of parameters are determined. The results of our investigations on the complex dynamics appearing in the system, including spatiotemporal chaos, intermittency, and turbulence are contained in Sect. IV.  Our concluding remarks are summarized in Section VI.}

\section{Experimental setup and Modelization}
Our experimental setup based on a coherently driven passive optical fibre ring cavity is depicted in Fig.~\ref{fig:ExperimentalSteup} . It is based on a resonant passive fibre ring cavity mainly made of 26.5~m-long segment of highly nonlinear optical fibre. 
The fibre combines a low group-velocity dispersion with a high nonlinear coefficient so as to 
enhance MI gain ($\beta_2 = -0.89$~ps$^2$ km$^{-1}$, {$\gamma = 10$~W$^{-1}$ km$^{-1}$} at 1552.4~nm). 
It also exhibits a low third-order dispersion ($\beta_3 \simeq 0.01$~ps$^3$ km$^{-1}$) that can be neglected 
(the zero dispersion wavelength is located below 1500~nm far from the pump wavelength). 
A 90/10 input coupler is used to close the fibre loop cavity, whereas a 99/1 output coupler permits to 
extract and analyse the intracavity field. The two couplers are made of SMF28 fibre 
with a total length of 1.5 m that belongs to the cavity (i.e., the total cavity length is equal to 28 m). 
The SMF28 fibre exhibits the following usual parameters: $\beta_2 = -21.7$~ps$^2$ km$^{-1}$ 
and $\gamma = 1.2$~W$^{-1}$ km$^{-1}$.  
 We split a continuous-wave (CW) laser at 1552.4~nm (linewidth <1 KHz) into two parts. 
 The first part, the {\em control wave}, is injected into the cavity to stabilize and fix the linear detuning 
 at the pump frequency $\omega_0$, with the help of a PID controller that finely tunes the laser wavelength. 
 The second part, the {\em pump wave}, is intensity modulated to generate 2~ns square pulses at 7.36~MHz 
 repetition rate (corresponding to the cavity's FSR). 
 This stage simultaneously enables the increase of the pump peak powers and the circumvention of
Brillouin backscattering within the cavity. The resulting quasi-CW pump is amplified by an erbium-doped 
fibre amplifier (EDFA), and launched into the cavity through the 90/10 coupler. 
 To minimize their mutual interaction, the control and pump beams are counter-propagating. Input polarization states are controlled via polarization controllers to excite a neutral axis of the cavity fibre. Temporal and spectral characterizations of intracavity field are provided by an intensity autocorrelator (with a temporal resolution of 10~fs and a full time window limited to 80~ps) and a high-resolution (2.5~GHz) optical spectrum analyzer.
The light field circulating in the effective {fibre} ring cavity (yellow box of Fig.~\ref{fig:ExperimentalSteup}) can be described by the following set of equations coupling the successive round-trip propagation described by the modified nonlinear Schr\"odinger equation (\ref{eq:Ikedamap1}) to the synchronously coherent injection { (\ref{eq:Ikedamap2})} 
 \cite{Coen1997,Tlidi2007,Haelterman1992}: 
 \begin{subequations} 
 \label{eq:Ikedamap}
\begin{eqnarray}
\partial_z A_{m}\left(z,T\right) &=&-\alpha_f A_{m}\left(z,T\right) + i\sum_{n\geqslant 2}\frac{i^n\beta_n}{n!}\partial_T^nA_{m}\left(z,T\right)\nonumber\\
&&+i\gamma A_{m}\left(z,T\right)\int_0^\infty R\left(T^\prime\right)\left|A_{m}\left(z,T-T^\prime\right)\right|^2dT^\prime 
\label{eq:Ikedamap1}\qquad\\
A_{m+1}\left(0,T\right)&=& \sqrt{\theta} E_i\left(T\right)+\sqrt{\rho} A_{m}\left(L,T\right)e^{-i\Phi_0}.
\label{eq:Ikedamap2}
\end{eqnarray}
 \end{subequations} 
Here $t_R$ stands for the round-trip time which is the time taken by the pulse to propagate along 
the cavity with the group velocity, $\Phi_0$ is the linear phase shift, $\theta\ \left(\rho\right)$ 
is the mirror transmission (reflection) coefficient, and $L$ is the cavity length. 
{ The complex envelope of the electric field inside the cavity at the $m$-th round trip is $A_m$}. Each of the coefficients $\beta_n$ is responsible for 
the $n$-th order dispersion, $\gamma$ is the nonlinear coefficient, and $\alpha_f$ is the attenuation along the fiber. 
The independent variable $z$ refers to the longitudinal coordinate while $T$ is the time in a reference frame 
moving with the group velocity of the light, and $R(T)$ is the nonlinear response including both instantaneous 
(Kerr effect) and delayed contributions (Raman effect). 
From transmission measurements with a  continuous wave (CW) pump the cavity at 1552.4 nm (linewidth <1 kHz) 
the finesse $\mathcal{F}=2\pi/\theta_{\mathrm{eff}}$ is nearly 19. 
Hence, the total power losses of the cavity $\theta_{\mathrm{eff}}$ including fiber absorption 
and coupler losses are about 30\% ($\theta_{\mathrm{eff}}=0.3$). 
From the fiber parameters, we obtain that our cavity is equivalent to a unique fiber ring cavity with 
$\beta_{2\mathrm{eff}}=-2$~ps$^2$km$^{-1}$ and $\gamma_{\mathrm{eff}}=9.6$~W$^{-1}$ km$^{-1}$. 
Therefore, without
loss of generality,  the evolution of the electric field inside the cavity is well described by the Lugiato-Lefever equation (LL model) \cite{Lugiato1987,Haelterman1992}:
\begin{equation}
\frac{\partial \psi}{\partial t}=S-(1+i\Delta)\psi-i\eta\frac{\partial^{2} \psi}{\partial \tau^{2}}+i\vert\psi\vert^{2}\psi,
\label{eq:LL}
\end{equation}
where  $\alpha=\theta_{\mathrm{eff}}/2$, $S=2E_i\sqrt{\gamma L}$, $\psi=A_m\sqrt{\gamma L/\alpha}$, { $t=\alpha t^\prime/t_R$}, 
and $\tau=T/T_n$ { with $T_n=\sqrt{\left|\beta_2L\right|/(2\alpha)}$}. $\Delta=(2k\pi-\Phi_0)/\alpha$ is the detuning with 
respect to the nearest cavity resonance $k$.  The coefficient $\eta=\pm 1$ is the sign of the group velocity dispersion 
term and $T_n=\sqrt{\left|\beta_2L\right|/(2\alpha)}$ and { $t^\prime\equiv mt_R$ { accounts for} 
the continuous time introduced to account for the evolution of the intracavity field over the round trips}.

 { In the bistable regime, the latter equation can exhibit soliton-like solutions (i.e., cavity solitons) \cite{Leo2010,leo2013a,herr2014}. Increasing the driving {strength} this soliton first undergoes an Andronov-Hopf instability yielding to self-pulsating localized state. For larger values of the pump intensity, this oscillatory localized state, in turn, becomes unstable. The evolution of the resulting complex state has been demonstrated 
 to be  { of} a spatiotemporal {chaotic nature} \cite{Liu2017a}. In this region of parameters, some of the emerging erratic spatiotemporal localized states have been statistically shown to satisfy the criteria of extreme events \cite{Coillet2014}. 
 From a general point of view, the extensive literature in the highly nonlinear bistable regime contrasts with those of the monostable case 
 {that we are interested in.}  The reason is twofold, (i) bistability makes more accessible the experimental observation of solitons, 
(ii) the { complex} nonlinear dynamics is believed { richer} than in the monostable regime. Indeed, { in the monostable regime, most} studies are dominated by the weakly nonlinear analysis. In that case,  { only MI is present and the system remains stable and} evolves in a regular way \cite{Lugiato1987,Braje2009,Leo2013,Godey2014,Bendahmane2017,Liu2017b}. The corresponding periodic pattern is characterized by a triangular comb spectrum \cite{Braje2009,Leo2013,Godey2014,Bendahmane2017,Liu2017b} whose behavior in the highly nonlinear regime, however, remained unexplored.

To fill this lack, we} have then performed a set of spectral and temporal measurements for a normalized linear detuning set to $\Delta =2(2\pi k-\phi_0)/\theta_{\mathrm{eff}} = 0.55$. Here $k$ refers to the nearest resonance and $\theta_{\mathrm{eff}}/2=\alpha=\pi/\mathcal{F}$ represents the power lost per roundtrip, with $\mathcal{F}$ the effective {finesse} of the cavity. For this value of the detuning, the setup is said to be in the monostable regime where no cavity soliton or complex dynamics related to bistability can take place, which prevents the formation of dissipative Kerr solitons. Only the Turing pattern or modulation instability (MI) can be observed at the onset of the cavity emission.
{
 
Figure~\ref{fig2:expspectra} depicts the evolution of the intracavity spectrum while tuning the input peak power. The first sub-figure
Fig.~\ref{fig2:expspectra}(a) represents the spectrum recorded for an input peak power $P_0 = \left|A_{in}\right|^2 = 0.16$~W (slightly above the MI threshold, $P_{th}=0.15$~W). Two weak MI sidebands detuned by 407~GHz appears in the spectrum. By increasing the pump power, we observe their amplification and frequency detuning (Fig.~\ref{fig2:expspectra}(b)). Also, the dynamics of cascade of MI sidebands appears progressively forming a triangular Kerr frequency comb (in a log-scale) composed by four harmonics of the MI frequency, equally spaced and detuned from the pump by 530~GHz (Fig.~\ref{fig2:expspectra}(c)). For higher pumping power (Fig.~\ref{fig2:expspectra}(d)), intermediary spectral peaks centered between the MI bands start to growth. 
\begin{figure}
\begin{center}
 \includegraphics[width=.45\textwidth]{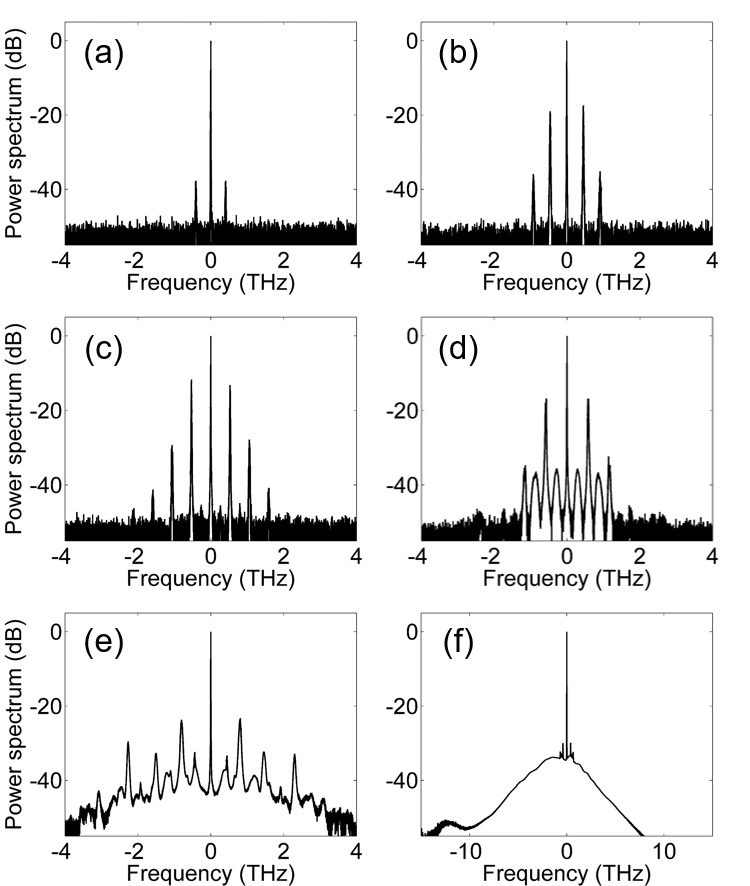} 
\caption{Experimental results. Intracavity spectra recorded for different input pump powers: (a) 0.16~W, (b) 0.24~W, (c)
0.56~W, (d) 0.9~W, (e) 3.1~W and (f) 20.7~W.}
\label{fig2:expspectra}
\end{center}
\end{figure}
\begin{figure}
\begin{center}
 \includegraphics[width=.45\textwidth]{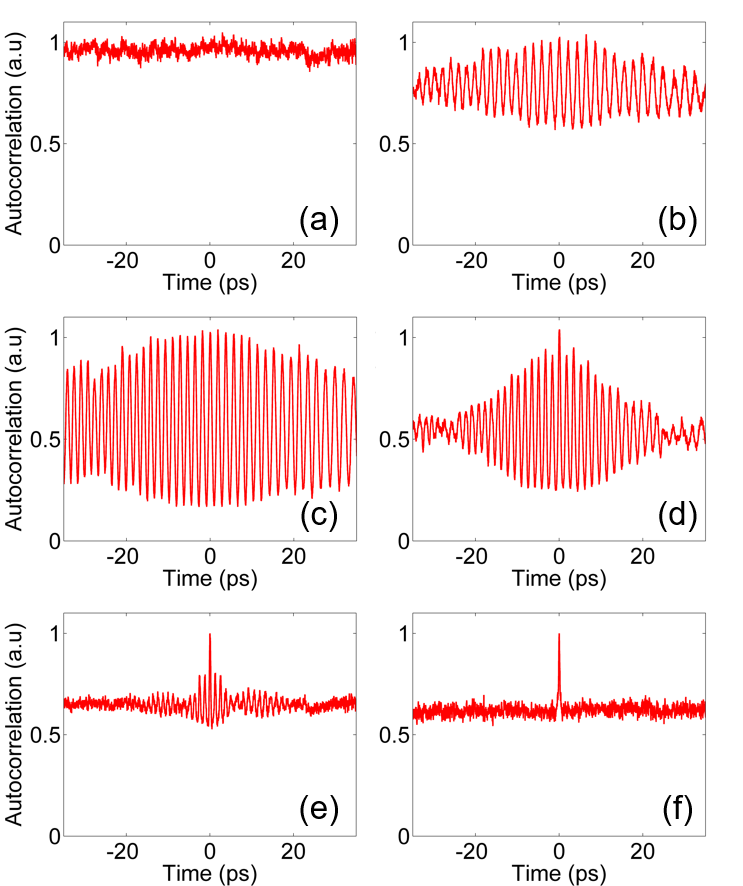} 
\caption{Experimental results. Autocorrelation traces recorded for different input pump powers: (a) 0.16~W, (b) 0.24~W,
(c) 0.56~W, (d) 0.9~W, (e) 3.1~W and (f) 20.7~W.}
\label{fig3:expautocorr}
\end{center}
\end{figure}

While increasing further the pump power, a more complex and broader spectrum develops (Fig.~\ref{fig2:expspectra}(e)). For $P_0 = 20$~W, a broad continuous triangular spectrum is formed (Fig.~\ref{fig2:expspectra}(f)). We also observe the emergence of the broad Raman Stokes component detuned by $-13$~THz from the pump. We cannot further increase the power because of the formation of a
Brillouin wave inside the cavity which destabilizes the locking loop.
\begin{figure}
\begin{center}
\includegraphics[width=.45\textwidth]{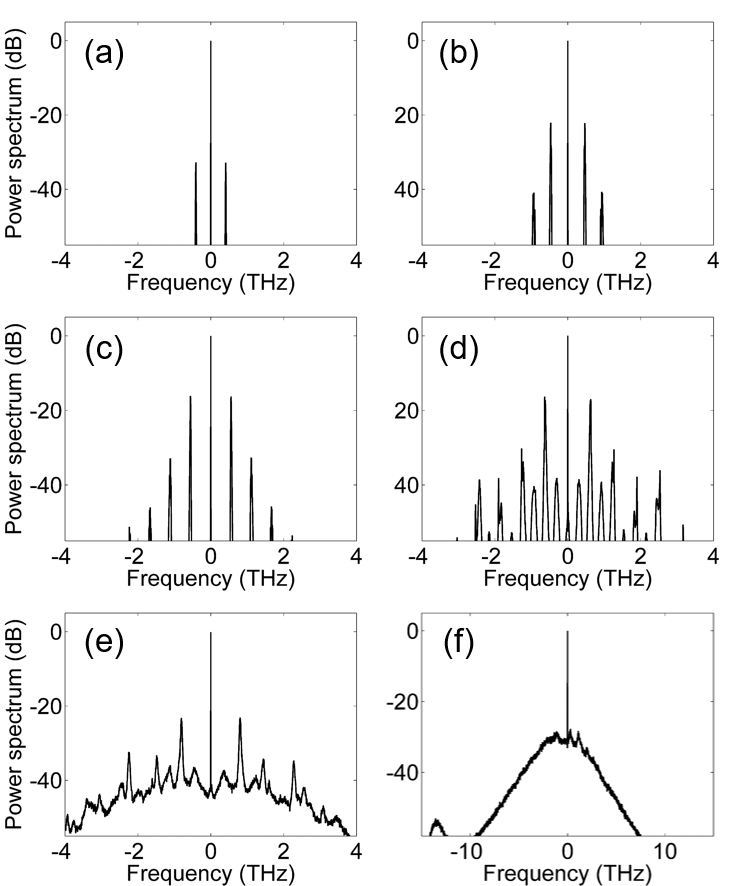} 
\caption{Numerical results  obtained from Eqs.~(\ref{eq:Ikedamap}). Intracavity spectra obtained for different input pump powers: (a) 0.188~W, (b) 0.24~W, (c)
0.45 W, (d) 0.9 W, (e) 3.1~W and (f) 12.5~W.}
\label{fig4:ikedaspectra}
\end{center}
\end{figure}

{Moreover,} we performed temporal measurements of the intracavity field by means of a background-free second-harmonic 
autocorrelator. Figure~\ref{fig3:expautocorr} displays the distinct recorded autocorrelation traces corresponding to the spectra from Fig.~\ref{fig2:expspectra}.
{In} the beginning, for a power just upon the MI threshold, we recorded a nearly flat trace corresponding to a quasi-continuous intensity field (Fig.~\ref{fig3:expautocorr}(a)). With the appearance and growth of MI sidebands, the trace becomes regularly modulated (Fig.~\ref{fig3:expautocorr}(b)). The contrast of this modulation is maximum when the triangular comb is reached (see Fig.~\ref{fig3:expautocorr}(c)) indicating the formation of a short-pulse pattern on a finite background inside the cavity. The period of the pulse train is about 1.85~ps in agreement with the expected 530~GHz repetition rate driven by the MI frequency. The central peak corresponds to the autocorrelation of a single pulse, whereas the adjacent peaks are the cross-correlations between neighboring pulses.
With the growth of the additional spectral peaks between the MI harmonics, the contrast seen in the trace starts to be degraded (Fig.~\ref{fig3:expautocorr}(d)). Since the cross-correlations between neighboring pulses is highly sensitive to neighboring pulse differences, we can infer the degradation of the formed pulses train because of these new sidebands. For high pump powers, the modulation completely disappears and only a central peak of coherence remains (Fig.~\ref{fig3:expautocorr}(e-f)). This reveals the strong intensity incoherence of the intracavity field (over a single round-trip). 
 \begin{figure}
\begin{center}
\includegraphics[width=.45\textwidth]{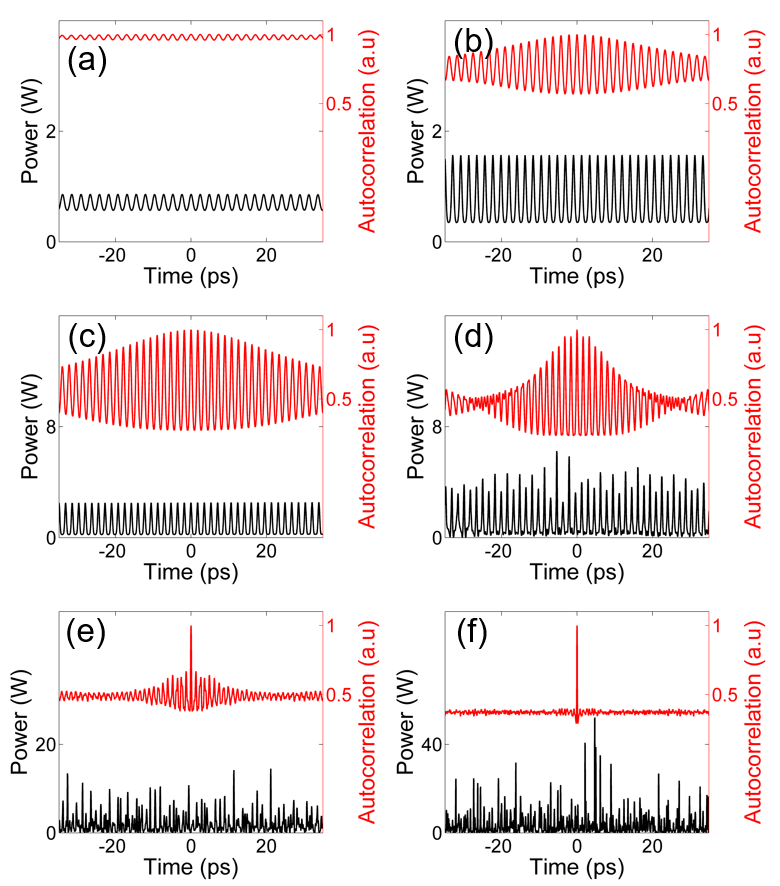} 
\caption{Numerical results  obtained from Eqs.~(\ref{eq:Ikedamap}). Intracavity temporal profiles and corresponding autocorrelation signals obtained for different pump powers: (a) 0.188~W, (b) 0.24~W, (c) 0.45 W, (d) 0.9 W, (e) 3.1~W and (f) 12.5~W.}
\label{fig5:ikedaautocorr}
\end{center}
\end{figure}
\begin{figure}
\begin{center}
 \includegraphics[width=.45\textwidth]{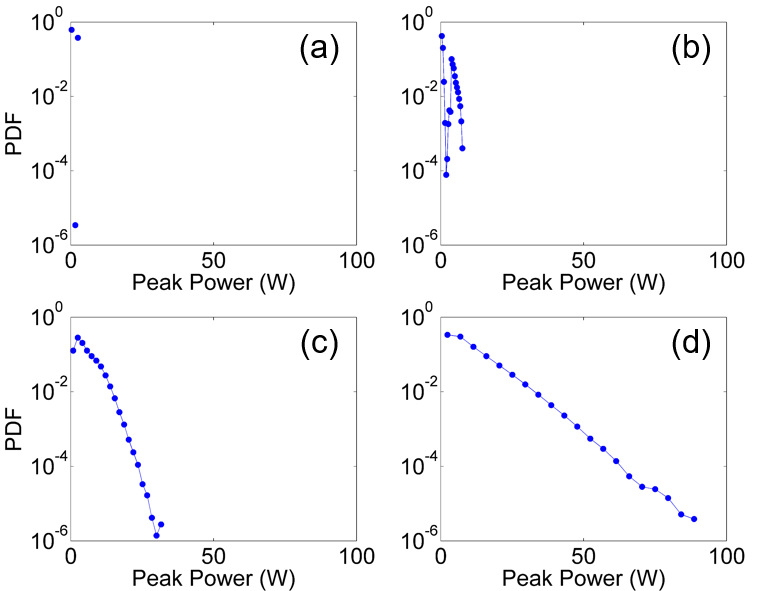} 
\caption{Numerical results obtained from Eqs.~(\ref{eq:Ikedamap}). Probability density function of the intensity peaks of the circulating intracavity field calculated for different pump powers: (a) 0.45~W, (b) 0.9W, (c) 3.1W, and (d) 12.5~W.}
\label{fig6:ikedapdf}
\end{center}
\end{figure}

{ We have also performed  numerical simulations based on Eqs.~(\ref{eq:Ikedamap}).} 
Figure~\ref{fig4:ikedaspectra} shows the different simulated intracavity spectra obtained after 10,000 cavity round trips ({i.e.,} 
a stationary state circulating in the cavity) as a function of the input peak power. The numerical results are found to be in {a} good 
qualitative agreement with experiments. At first, MI sidebands and their harmonics appear and get amplified while increasing the pump power (Fig.~\ref{fig4:ikedaspectra}(a-c)). 
Then, intermediary sidebands start {to grow} and the spectrum broadens progressively until a continuous triangular shape 
is reached (Fig.~\ref{fig5:ikedaautocorr}(f)). Temporal analysis of the simulated intracavity field was also carried out (see Fig.~\ref{fig5:ikedaautocorr}). 
In particular, the autocorrelation traces (red lines, right axis) are found to be similar to the experimental ones. \ Fig.~\ref{fig5:ikedaautocorr} also shows 
the corresponding field evolution (black lines, left axis). 
Details from the field evolution {allow confirming} the experimental results. At first, the appearance of the MI sidebands {induces} 
the modulation of the intracavity field until the formation of a train of short pulses (of about 360~fs) on a finite background (Fig.~\ref{fig5:ikedaautocorr}(a-c)). 

Next, the growth of intermediary bands progressively degrades the regular pulse train, first, by doubling the period (Fig.~\ref{fig5:ikedaautocorr}(d)) 
and then by strongly decreasing the intensity coherence, {i.e.,} an irregular wave forms with the appearance of some rare pulses with high peak powers compared to the average. To characterize this behavior from a statistical point of view, we computed the probability density distribution of the intensity peaks of the circulating intracavity field (using data from the last 1,000 roundtrips). Fig.~\ref{fig6:ikedapdf} displays the probability density function versus the pulse peak power. We remark that for an input power of $\sim1$~W (or below), we mostly observe a distribution centered on a unique value of peak power that evolves towards a bimodal distribution. This behavior is directly related to the initial regular pulse train 
on a finite background whose period becomes doubled due to the growth of intermediary spectral bands (Fig.~\ref{fig4:ikedaspectra}(d)). 
After {that,} the temporal pattern is {wholly} degraded (Fig.~\ref{fig5:ikedaautocorr}(d{}-e)), the corresponding statistics broadens and becomes progressively more tailed (i.e., towards a right-skewed distribution, see Fig.~\ref{fig6:ikedapdf}(c)). For further increased power, the statistics develops a long tail indicating the presence of rare events with extremely high peak powers (Fig.~\ref{fig6:ikedapdf}(d)). These events can reach a peak power ten times higher than the average intracavity power. Note that the calculated rogue  wave intensity  threshold, defined as $I_{RW} = 2 I_{SI}$ ($I_{SI}$ being the significant intensity equal to the mean of the upper third of events in the distribution), is $I_{RW} =$ 5.1, 9.2, 17.6, and 34.9, respectively for subplots from Fig.~\ref{fig6:ikedapdf}. As a consequence, this criterion confirms that extreme events then appear for input power powers $\sim2$~W.
{ To the best of our knowledge, extreme events have never been reported in this monostable regime of a passive Kerr-type cavity. For a better understanding of the mechanism behind this prediction, the rise of the observed spatiotemporal complexity needs to be deeply characterized. However, it is an elusive task to apply many of tools of dynamical systems on Eqs.~(\ref{eq:Ikedamap}). Hence, for the sake of simplicity and without loss of generality, in what follows, we will use the reduced equation (\ref{eq:LL}). For the accuracy of our prediction based on this model (\ref{eq:LL}), it is worthy to determine the effective parameter after the experimental measurements.} 

Close to the emission threshold}, from Eq.~(\ref{eq:LL}), the dimensionless frequency shift of the MI is given by $\Omega_{c}=\sqrt{2I_s-\Delta}/\left(2\pi\right)$ when $\eta=-1$. $\Omega_{c}^{2}$ is then a linear function of the intracavity field with the $y$-intercept corresponding to $-\Delta$. By setting $P_{in}$ the mean intracavity intensity and  $\omega_{c}$ the physical frequency shift we can infer that $\omega_{c}^{2}=aP_{in}+b$ with $a$ and $b$ to be determined.  $P_{in}$ can be also given in ratio of its value at the MI threshold $P_{in}^{th}$, such that $I_{s}=P_{in}/P_{in}^{th}$. Hence, it follows: 
\[
\omega_{c}^{2}=\frac{aP_{in}^{th}}{2}\left(2I_{s}+\frac{2b}{aP_{in}^{th}}\right).
\]
 Consequently,
\begin{eqnarray*}
\Delta & = & -\frac{2b}{aP_{in}^{th}}, \ \mathrm{and}\\
T_{n} & = & \frac{1}{2\pi}\sqrt{\frac{2}{aP_{in}^{th}}}=\sqrt{\frac{\left|\beta_{2}L\right|}{2\alpha}}.
\end{eqnarray*} 

\begin{figure}
\begin{center}
\includegraphics[width=.5\textwidth]{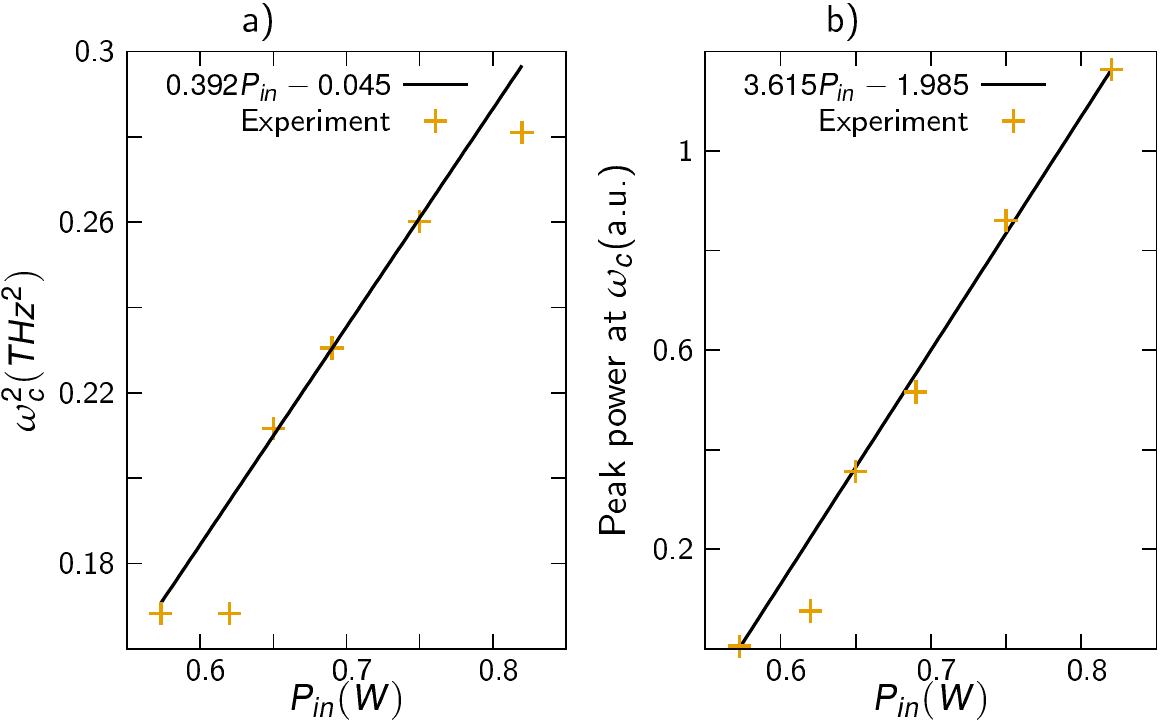}
\caption{Interpolation of experimental data extracted from the spectrum. The values of these figures have been extracted from the experimental measured  spectra.\label{fig:exp_data_sampling}}
\end{center}
\end{figure}

On the other hand, in the weakly nonlinear regime, the amplitude of the MI varies linearly with the distance at the threshold. 
From the experimental {data,  we have extracted} the $\omega_{c}$ and the corresponding magnitude. The result is given in Fig.~\ref{fig:exp_data_sampling}.
From Fig.\ref{fig:exp_data_sampling}b the threshold is obtained as $P_{in}^{th}=1.985/3.615=0.55$~W. By considering the effective length $L=28$~m and group velocity dispersion parameter $\beta_{2\mathrm{eff}}=-2$~ps$^2$km$^{-1}$ we have calculated the effective parameters corresponding to the configuration of the setup. The most relevant are shown in Table~\ref{tab:params_validation}. 
\begin{table}[H]
\begin{center}
\begin{tabular}{cccccc}
\cline{2-6} 
 & Calculated &  &  &  & Expected\tabularnewline
\hline 
\hline 
$\Delta$ & 0.42 &  &  &  & 0.55\tabularnewline
\hline 
$T_{n}$(ps) & 0.48 &  &  &  & 0.43\tabularnewline
\hline 
$\alpha=\left|\beta_{2}L\right|/\left(2T_{n}^{2}\right)$ & 0.12 &  &  &  & 0.15\tabularnewline
\hline 
$\mathcal{F}=\pi/\alpha$ & 26.36 &  &  &  & 20.94\tabularnewline
\hline 
$\theta_{\mathrm{eff}}=2\alpha$ & 0.24 &  &  &  & 0.30.\tabularnewline
\hline 
\end{tabular}
\end{center}
\caption{Comparision between extracted and expected effective parameters.}
\label{tab:params_validation}
\end{table}

\begin{figure*}
\centering 
\includegraphics[width=.75\textwidth]{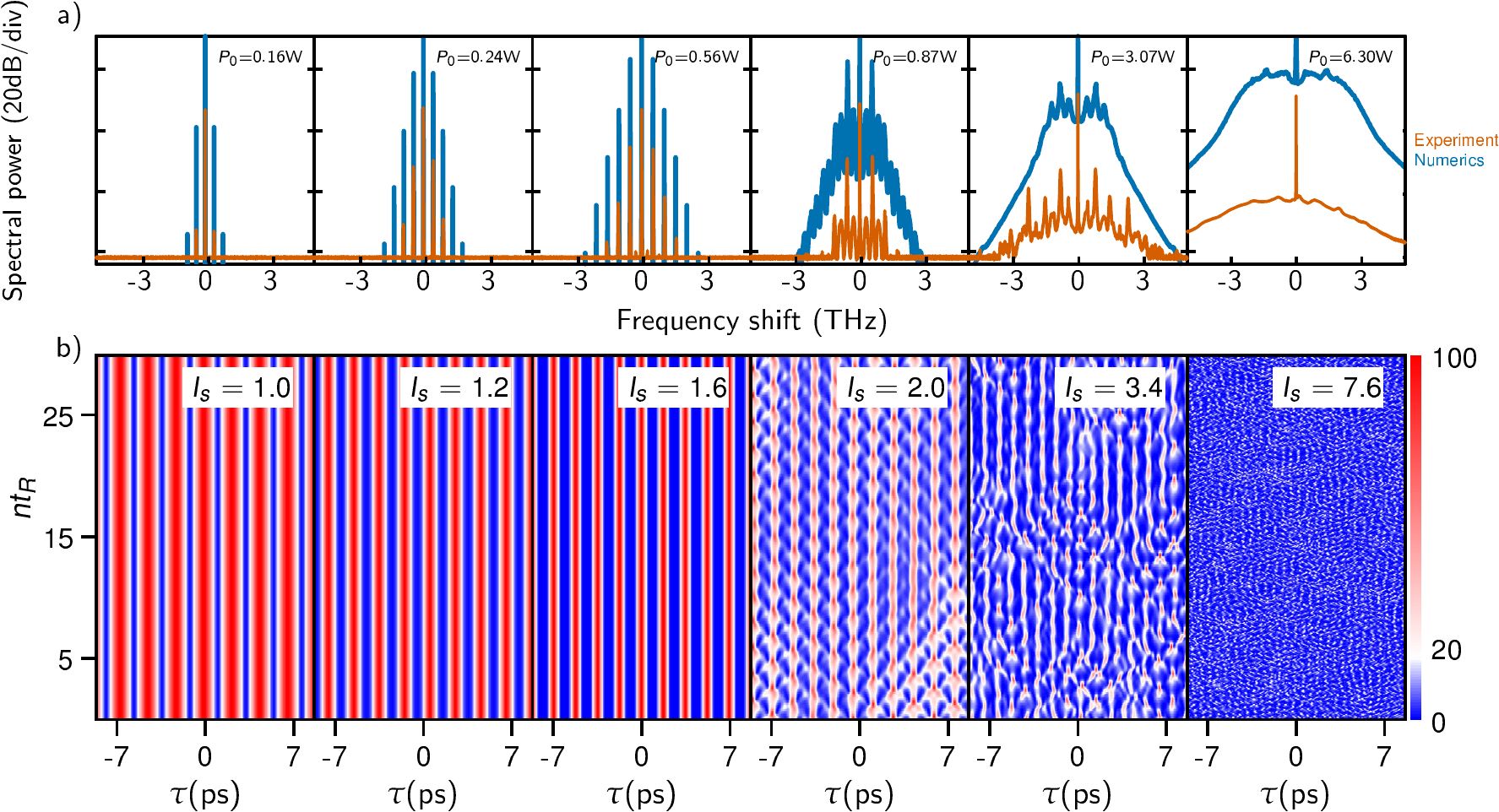}
 \caption{(a)Experimental (red) and numerical (blue) intracavity spectra  recorded for different input pump powers. For clarity, experimental spectra are vertically offset by -20dB.  (b) Evolution of the intra-cavity intensity profile over several roundtrips, computed numerically from the Lugiato-Lefever equation {(\ref{eq:LL})} for the same values of the pump powers in the top panel. The value of $I_s$ is calculated in ratio of the intracavity mean power at the threshold {$P^{th}_{in}$}. The detuning parameter is $\Delta=0.55$.}
 \label{fig:numerical_xt_diagrams}
 \end{figure*}
 
\section{From modulation instability to spatiotemporal chaos}
\begin{figure*}
\centering 
\includegraphics[width=\textwidth]{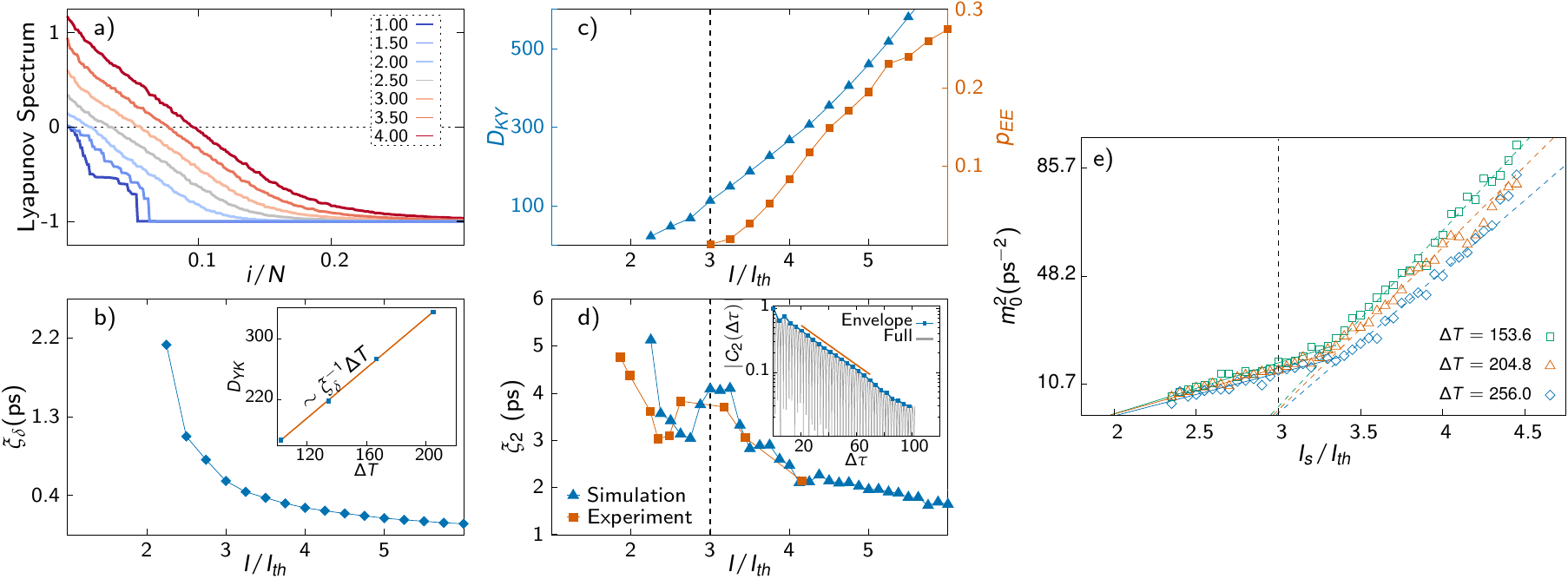}
\caption{ (a) Lyapunov spectra computed from the numerical integration {of} Eq.~(\ref{eq:LL}) { with} increasing  pump power. (b) The Lyapunov dimension density dimension. (c) Dependence of the Kaplan-Yorke dimension (blue triangles) and the proportion of extreme events (red squares) on the pump power. (d) The equal-time two-points correlation length with respect to the input parameter. (e) Square of the maximum exponential decay  rate of the laminar periods inside { a roundtrip of} temporal profiles.} 
\label{fig:CharacteristicLength} 
\end{figure*}

After the validation of expected parameters, we now 
follow the evolution of the modulation instability from the weak to the unexplored large driving strength regime.
Figure~\ref{fig:numerical_xt_diagrams}a shows the evolution of the recorded intracavity power spectrum when the power of the driving beam (i.e., the pump) increases. {We first recall the two} significant { and striking} features that have drawn our attention to the evolution of these spectra. The first one is a transition from a 
purely triangular comb (in {a} log. scale) to a quasi-periodic-like spectrum formed by the appearance of secondary instability bands in between the primary comb teeth. The second change arises while a continuous spectrum is observed in place of the comb-like profile.  Later, this continuous spectrum only broadens as the pump power increases.  
A typical behavior { is} shown in Fig.~\ref{fig:numerical_xt_diagrams}a obtained from the numerical simulations based on the Lugiato-Lefever equation (\ref{eq:LL}). The bottom panel (Fig.~\ref{fig:numerical_xt_diagrams}b) shows the corresponding density plot of the temporal traces for successive round trips. The key feature of this evolution is the apparent complexity of the dynamics when { the spectrum goes beyond the triangular-shape comb} ($I_s\simeq 2.0$). The analysis of the structural changes in the optical spectrum is performed through the computation of the Lyapunov spectrum (see Appendix~\ref{app:lyap_spec} for details) of this reduced model for an increasing pump power.  The results are displayed in Fig.~\ref{fig:CharacteristicLength}a.  { From this figure, we can observe a value of the pump power ($I_s=2.0$) below which the continuous Lyapunov spectra are composed of only negative exponents. Above this value, the spectra show a region with positive exponents that increases with pump power suggesti{ng} the emergence of a chaotic dynamics in the spatiotemporal complexity observed for $I_s>2.0$ in Fig.~\ref{fig:numerical_xt_diagrams}b. An indicative quantity of the number of degree of freedom needed to describe { the} chaotic dynamics is the dimension of the strange attractor. An estimator { of} this dimension, based on the Lyapunov spectrum, has been conjectured by Kaplan and Yorke \cite{Kaplan1979}. The dependance of this Kaplan-Yorke dimension $D_{KY}$ on the control parameter is shown by Fig.~\ref{fig:CharacteristicLength}c, which confirms the increasing of the complexity with the pump power. { Besides} this growth of the Kaplan-Yorke dimension with the pump power, we must also consider the conjecture that the dimension increase{s} with the {\em size} of the system--expansive nature of the spatiotemporal chaos \cite{Clerc2013,Manneville1995,Ott2002,Abarbanel2012,Ruelle1982}. To focus only on the effect of the pump power, it is necessary to use an intensive measurement of the complexity. This can be done through the Lyapunov dimension density (see Appendix~\ref{app:lyap_spec} and \ref{app:lyap_dens_dim}). Indeed, $D_{KY}$
may change linearly with the volume of the system \cite{Ruelle1982,Cross1993}. That is, for a 1D system, $D_{KY}=\xi_\delta^{-1} \Delta T$ 
where $\Delta T$ is the extension of the system and $\xi_\delta$ represents the { Lyapunov dimension density} of the system for a {fixed} value of the pump power. Therefore{,} it is an intensive quantity that provides an estimation of the extension of the independent subsystems generated by the chaotic dynamics. We have performed a systematic calculation of the Lyapunov dimension density $\xi_\delta$ and the results are given in Fig.~\ref{fig:CharacteristicLength}b. It appears  that the  the characteristic range of chaotic fluctuations  decreases as the driving power increases.} 

Considering that the {\em spatiotemporal} chaos produces { a} subsystem of {the} order of $\xi_\delta$, one might wonder how far these subsystems interact. 
This can be addressed by characterizing the average temporal disorder provided by the {equal-time two point correlation length $\xi_2$ obtained from the exponential decay of the following integral
\cite{Ohern1996,Egolf1994,Cross1993}:
\begin{equation}
\label{eq:2ptcorr}
C\left(\Delta\tau\right)=\langle\left(\psi\left(\Delta\tau+\tau^\prime,t\right)-\langle\psi\rangle\right)\left(\psi\left(\tau^\prime,t\right)-\langle\psi\rangle\right)\rangle.
\end{equation}
The brackets $\langle\cdot\rangle$ stand for the average process (see Appendix~\ref{app:corr2_dim} for practical computation).} For our system, the result is shown in Fig.~\ref{fig:CharacteristicLength}d. 
We clearly observe a non-monotonic evolution of the correlation time in a subsequent range of pump power. 
This behavior is well confirmed by corresponding experimental data, { suggesting that} short-range fluctuations measured by $\xi_\delta$ may be decoupled from the long-range order measured by $\xi_2$\cite{Ohern1996}. { In general, for unidimensional (1D) { systems,} $\xi_\delta$ and $\xi_2$ are proportional, specifically when the later one is computed from the magnitude of the considered field. For now, only one study {has} reported different evolutions of  $\xi_\delta$ and $\xi_2$. However, in that case  $\xi_2$ has been computed from the phase \cite{Egolf1995}. Hence, the decoupling between $\xi_\delta$ and $\xi_2$ measured here from { the same dynamical component--the magnitude of the intracavity field}--is an unexpected observation.}

{ From Figs.~\ref{fig:CharacteristicLength} b) and c) we can see that the correlation length { $\xi_2$} can be larger up to ten times the Lyapunov density dimension. 
Therefore, a {significant} number of the subsystems may be dynamically correlated. Two {central} questions arise from this observation. Are the spatiotemporal chaotic subsystems contiguous or not and if not what is the characteristic time in between? An appropriate method to answer these questions is { to address the dynamics of} the different points of the system in terms of laminar (regular) and turbulent (irregular) periods according to a cut-off threshold \cite{Chate1987}.
 {
If the chaotic subsystems are not contiguous, the dynamics may lead to a fluctuating mixture of turbulent and laminar periods\cite{Ciliberto1988,Turitsyna2013}, the so-called spatiotemporal intermittency. If so, the probability distribution of the laminar periods should be a mixture of the power law 
({long-range} correlation) and the exponential law (short range)\cite{Ciliberto1988}. In this mixture distribution, the exponent of the power law $\left(\mu\right)$ is expected to be insensitive to the input parameter or the value of the cut-off threshold. By contrast, the decay rate $\left(m\right)$ of the exponential law depends on both parameters. Since the dependence on the cut-off threshold is only a decaying {exponential,} the  y-intercept $\left(m_0\right)$ contains the dependence on the control parameter $S$ through $I_s$. We will refer to $m_0$ as the intrinsic exponential decay.
 
 \section{Transition to amplitude turbulence via spatiotemporal intermittency	}
Before going further, it may be helpful to describe with details how we have detected the laminar and turbulent domain in the temporal traces. 
This characterization was done following the process explained in  Ref.~\cite{Ciliberto1988}. 
As reminded by authors {in the} latter reference, {\em the distinction between laminar and turbulent regions solely in terms of the amplitude does not have a rigorous justification.} 
However, as we will see later, the domain boundaries obtained using the amplitude also correspond to a local change in the periodicity of the patterns. Notice that a rigorous distinction condition should be based on local frequency fluctuations over a roundtrip.
\begin{figure*}
\begin{center}
\includegraphics[width=.75\textwidth]{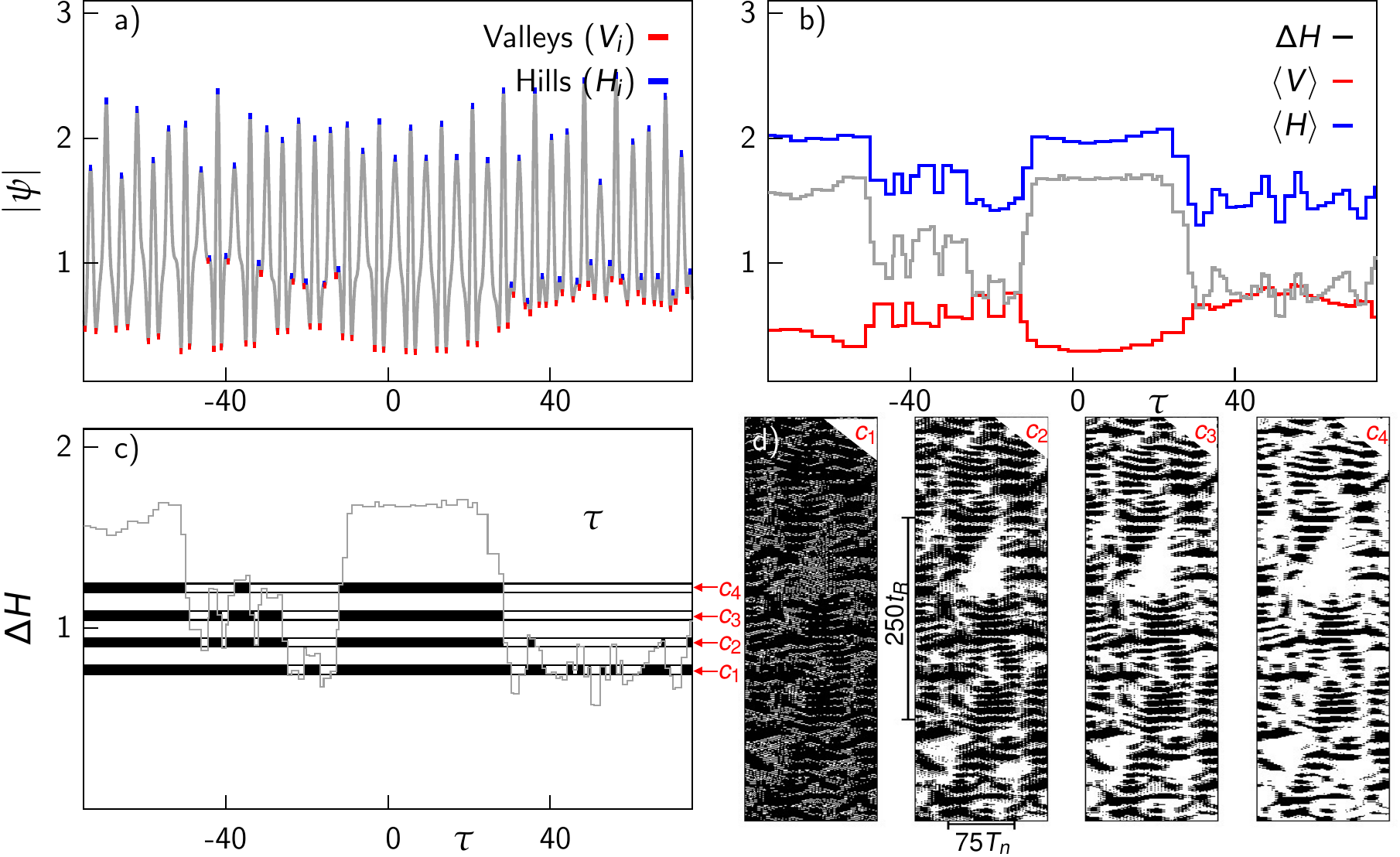}
\caption{Illustration of the process transforming the temporal trace into black (turbulent) and white (laminar) region. In (a) the hills and valley are detected. (b) The temporal trace is transformed into the value of local hills (blue), valley (red) and peak-to-peak (gray, $\Delta H=\langle H\rangle_i-\langle V\rangle_i$). These local values are given by the moving average over three consecutive values: $\langle V\rangle_i = \left(V_{i-1}+2V_i+V_{i+1}\right)/4$. (c) Setting $c$ a detection threshold, the temporal trace is locally transformed into laminar (below threshold--white) or turbulent (below threshold--black). Varying this threshold we can see that the statistics of the black or white region changes. (d) Example of binarized roundtrip dynamics increasing (from left to right) the detection threshold.}\label{fig:from_prof_lam}
\end{center}
\end{figure*}

According to the aforementioned process separating our temporal profile in a laminar and turbulent region is done as follows:
\begin{enumerate}
\item Detecting hills and valley in the temporal tracer (Fig.~\ref{fig:from_prof_lam}a)
\item Computing the amplitude of the local peak to peak amplitude (Fig.~\ref{fig:from_prof_lam}b)
\item Fixing a threshold below which a local peak is said laminar or turbulent otherwise (Fig.~\ref{fig:from_prof_lam}c and \ref{fig:from_prof_lam}d).
\end{enumerate}
After the last step of this process, the statistical study of the laminar duration is done. 
According to the theory, spatiotemporal intermittency is characterized by a {power-law} 
decay of the number of laminar time duration, while {an exponential decay governs a fully amplitude turbulent regime}. 
However, in case of Ising-type transition the probability density distribution of the laminar time should follow a mixture function as suggested by authors in \cite{Ciliberto1988}. Figure~\ref{fig:mu_fIs} shows the log-log plots of the probability density function of laminar region when increasing the pump power. As expected, the slope of the best-linear fit exhibits almost no dependence neither on the detection threshold nor the control parameter. Considering the mean value $\mu=4.8\pm0.4$, 
we were able { always} to find a portion of the statistics that satisfies the law PDF$\left(x\right)\propto x^{-\mu}$, characteristic of a spatiotemporal intermittent evolution. Elsewhere, there is also some portions of the statistics that may decay exponentially: PDF$\left(x\right)\propto e^{-mx}$ (see Fig.~\ref{fig:exp_pdf}a). However, this decay rate depends on {$c$,} and the control parameter $I_s$ such that $m\left(I_s,c\right)=m_0\left(I_s\right)e^{-c/c_0}$ as shown in Fig.~\ref{fig:exp_pdf}b. Finally, as it can be seen from Fig.~\ref{fig:CharacteristicLength}e $m_0\left(I_s\right)$ itself presents a power law increase on the control parameter.  Indeed, different size of the system show the same linear evolution of $m^2_0$ {upon} $I_s$. This figure also allows to identify two critical points--values of the control parameter $I_s$ from which $m_0$ scales as a power law. At the first transition which occurs at $I_s\simeq 2$, the distribution of laminar periods follows a power law.
\begin{figure*}
\begin{center}
\includegraphics[width=.75\textwidth]{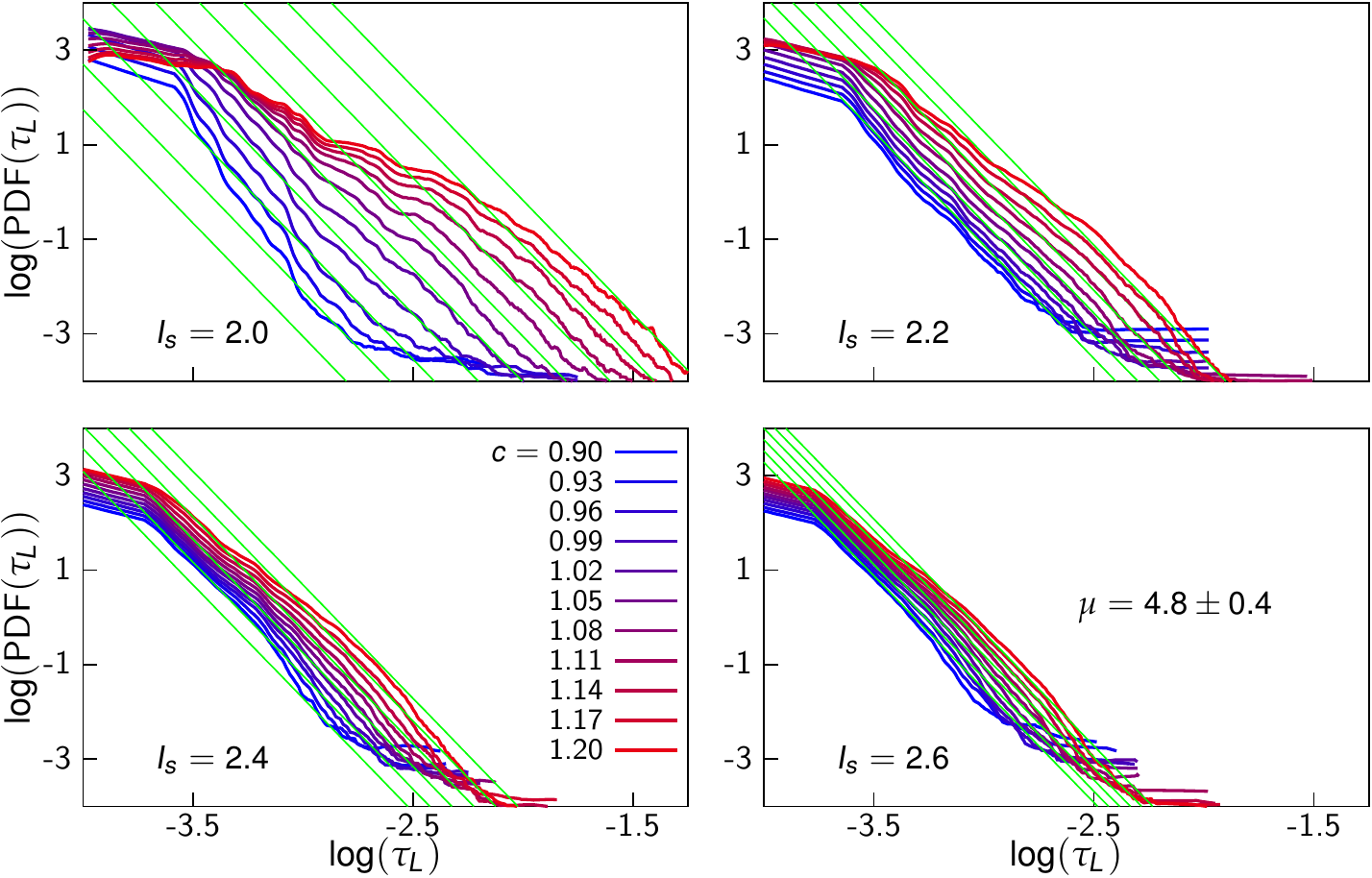}\caption{Log-Log plot of the probability density distribution of the laminar duration time ($\tau_L$) computed from the numerical simulations for $\Delta=0.5$, taking different values of the detection threshold $c$. The slope $\mu$ of the green lines correspond to the mean value of the best linear fit estimated with the PDF obtained for $2.\lesssim I_s\lesssim 2.75$. Therefore we can write that PDF$\left(x\right)=Ax^{-\mu}$.}\label{fig:mu_fIs}
\end{center}
\end{figure*}
\begin{figure*}
\begin{center}
\includegraphics[width=.75\textwidth]{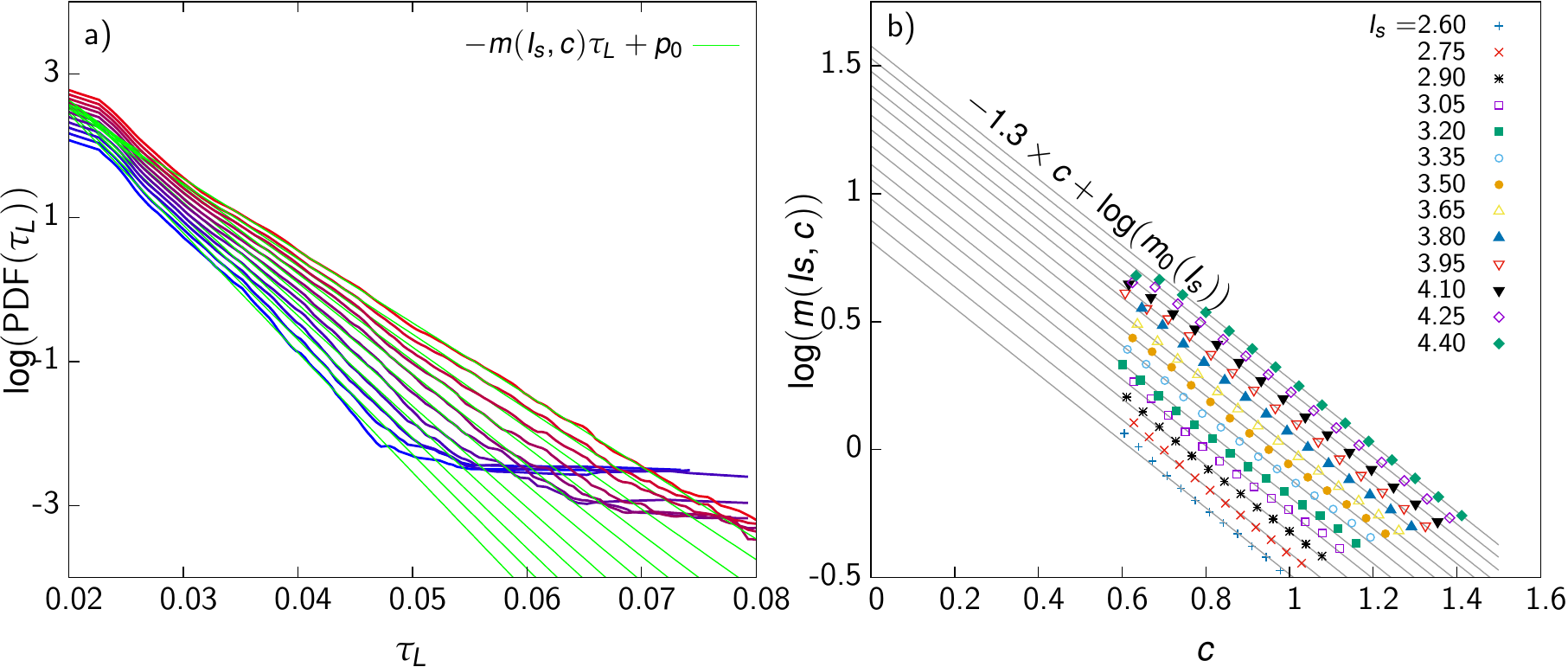}\caption{(a) semi-Log plot of the probability density distribution of the laminar duration time computed from the numerical simulations for $\Delta=0.5$, taking different values of the detection threshold $c$. For a given value of the control parameter, the slope $m$ of the green lines decreases as $c$ increases such that $m\left(I_s,c\right)=m_0\left(I_s\right)e^{-c/c_0}$. Hence, the intrinsic slope $m_0\left(I_s\right)$ corresponds to the $y$-intercept as illustrated by (b).} \label{fig:exp_pdf}
\end{center}
\end{figure*}
\begin{figure*}
\begin{center}
\includegraphics[width=.75\textwidth]{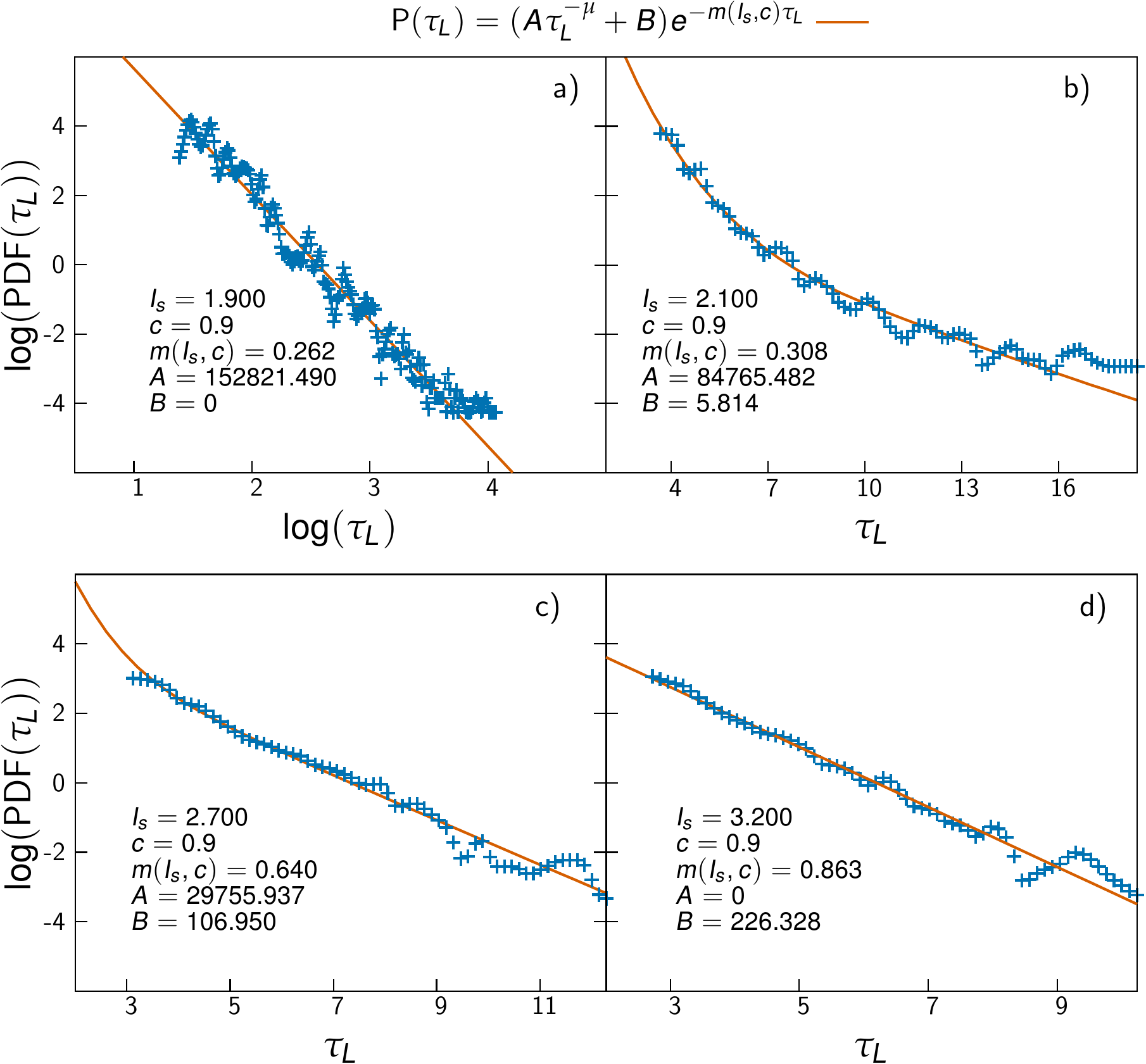}\caption{Probability density distribution (+) of the laminar duration time computed from the numerical simulations for $\Delta=0.5$, for $c=0.9$. The solid line corresponds to the fit following the function $P(x)=(Ax^{-\mu}+B)e^{-mx}$.} \label{fig:illust_pt}
\end{center}
\end{figure*}
}\begin{figure*}
\centering 
\includegraphics[width=1.\textwidth]{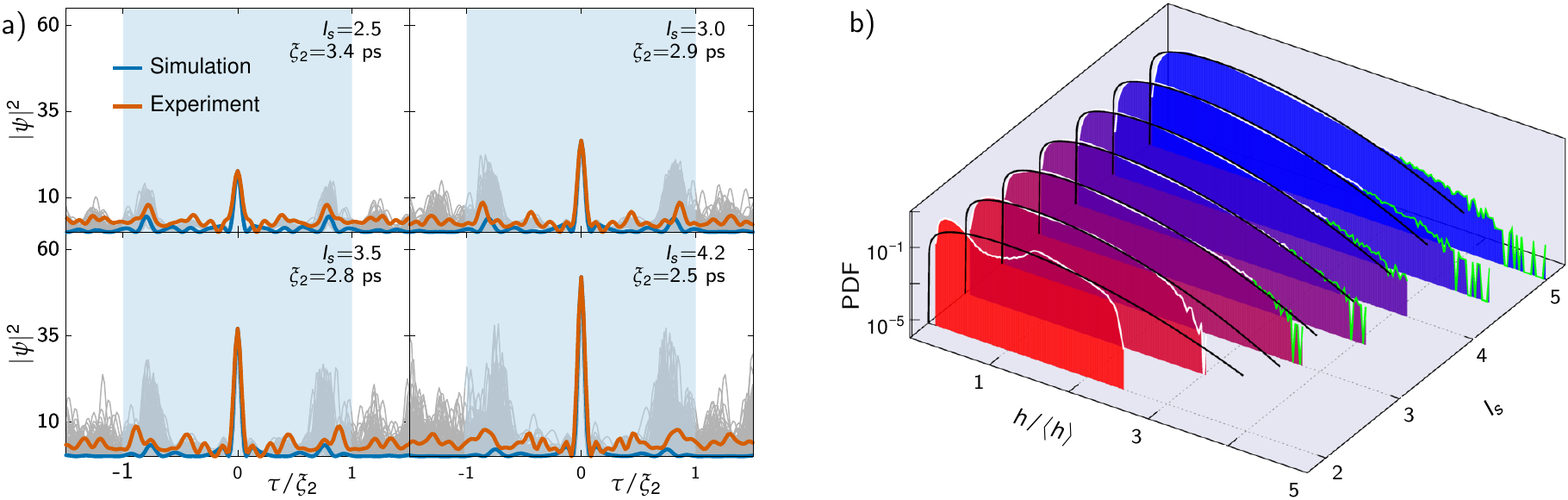}
\caption{(a) Superposition of the roundtrip temporal profiles containing an extreme event (gray lines). The blue (red) line corresponds to the function $\left|C\left(\Delta\tau\right)\right|^2$ obtained from the numerical (experimental spectra) data. Their maxima have been {rescaled} { to the value of the highest peak amplitude.} (b)Logarithmic scaled probability density functions of spatiotemporal peaks when increasing the pump power. The green line refers to peaks above twice the characteristic height and the black line is the Rayleigh distribution with unity mean value. } 
\label{fig:NumericalEEProfiles} 
\end{figure*}} 
Therefore, we conclude that $I_s\simeq 2$ {, that we set to  $I_s=I_s^{STI}$,} corresponds to the onset of the spatiotemporal intermittency. Since $I_s=I_s^{STI}\simeq 2$ also coincides with the onset of the spatiotemporal chaos, we can conclude that the spatiotemporal intermittency is the route to this chaos \cite{Chate1987,Ciliberto1988}. The second critical phenomenon which occurs at $I_s=I_s^{T}\simeq 3$ reveals the onset to the fully developed amplitude  turbulence \cite{chate1994spatiotemporal}, { given that the distribution of laminar periods follows an exponential law \cite{Chate1987}}. For $I_s^{STI}<I_s<I_s^{T}$ we were able to fit the probability distribution { as} $P(\tau)=(A\tau^{-\mu}+B)\exp(-m\tau)$ (see  Fig.~\ref{fig:illust_pt}a-d), typical of a phase transition process \cite{Ciliberto1988}. In this range of parameters, from the numerical data, we can observe the presence of intensity bursts in the intracavity field temporal profile. Figure~\ref{fig:NumericalEEProfiles}a shows the typical profile of 100 largest bursts when increasing the pump power.  It reveals that the largest peaks are always surrounded by smaller ones located at a position  { corresponding roughly to $\xi_2$}.  This evidences that $2\xi_2$ quantifies { the} region needed by the bursts to rise and disappear.  We have also plotted over these profiles the experimental two-points correlation function $\left|C\left(\Delta\tau\right)\right|^2$. We emphasize that the correlation function gives an accurate estimation of the temporal profiles of the intensity bursts, specifically inside the turbulent regime. We have then considered the statistical analysis of these intensity peaks in connection with extreme event studies {\cite{Kharif2003}. Figure~\ref{fig:CharacteristicLength}c shows the evolution of the ratio of bursts that can be considered as extreme events using the criterion defined in hydrodynamics\cite{Kharif2003} and Fig.~\ref{fig:NumericalEEProfiles}b gives some of the corresponding probability distributions. From these two results, we observe that the emergence of extreme events appears to coincide with the onset of the turbulent regime at $I_s=I_s^T$. Furthermore, in the vicinity of a local maximum of the correlation length and just before {the} emergence of the extreme events, we notice the bimodal shape of the {probability density function (PDF)}. This suggests the existence of two main sub-populations in the intensity profile like in the metal-insulator phase transition \cite{Landau1937,McLeod2017}. }
\section{Conclusion}
In {summary}, we have shown that the dynamics of the well-known periodic Turing patterns associated with the triangular-shape frequency comb in a Kerr-nonlinear ring cavity can be subject to transition to spatiotemporal chaos.  Our findings show that combining the study of different order parameters instead of trying to get the better one might be the sound approach to describe the spatiotemporal complexity. 
Once established that the whole system can be {split} into independent subsystems 
through the computation of the Lyapunov dimension length, the equal-time correlation length {has} provided the range over which their fluctuations are dynamically connected even not contiguous. Indeed, chaotic subsystems (turbulent) are shown to be separated by coherent subsystems (laminar). 
By analogy with {fluids,} the resulting mixture undergoes a transition to turbulence 
via spatiotemporal intermittency. Our findings also show that non-excitable passive resonators 
can generate spatiotemporal pulses whose probability distribution exhibits a long tail (i.e., extreme events).
Our results highlight how experiments in optics with controllable complexity using technology-driven 
components can be used to develop understanding of fundamental nonlinear dynamics in dissipative systems. We anticipate applications in establishing links between different branches of nonlinear science for which the occurrence of critical phenomena and extreme events is universal.

\section*{Acknowledgement}
SC, MT, AB, GM, BK gratefully acknowledge support from the French Agence Nationale de la Recherche trough OptiRoC project (Grant No. ANR-12-BS04-0001-011). BK had support from the French project PIA2/ISITE-BFC. M.G.C. thanks for the financial support of FONDECYT project {1180903} and {Millennium Institute for Research in Optics.}
S.C. and MT acknowledge the LABEX CEMPI (ANR-11-LABX-0007) as well as the Ministry of Higher Education and Research, Hauts de France council and European Regional Development Fund (ERDF) through the Contrat de Projets Etat-Region (CPER Photonics for Society P4S).

\appendix

\section{Lyapunov spectrum}
\label{app:lyap_spec}
Given a solution of the considered system, the Lyapunov spectrum characterizes the evolution of the perturbations around it. This spectrum is composed of the set of Lyapunov exponents. Positive exponents manifest a chaotic nature of the dynamics.

The procedure to obtain the Lyapunov spectrum is the following \cite{Skokos}. Let us take the linearized 
system $\partial_t\delta\textbf{X}=\textbf{J}\delta\textbf{X}$, where $\delta\textbf{X}$ is the perturbation 
around the considered solution and $\textbf{J}$ the respective Jacobian. Introducing a matrix $\textbf{L}$, that contains $n$ orthonormal vectors $\textbf{v}_i$
\begin{equation}
\textbf{L}\left(t= t_0\right)\equiv\left[\textbf{v}_1\quad \textbf{v}_2\quad\dots \quad\textbf{v}_n\right]=\begin{bmatrix}
    x_{11}       & x_{12} & x_{13} & \dots & x_{1n} \\
    x_{21}       & x_{22} & x_{23} & \dots & x_{2n} \\
    \hdotsfor{5} \\
    x_{d1}       & x_{d2} & x_{d3} & \dots & x_{dn}
\end{bmatrix},
\end{equation}
where $d$ is the dimension of the system and $n$ the number of Lyapunov exponents to be computed.  After a time increment $dt$, the matrix $\textbf{L}$ evolves to $\textbf{L}\left(t_0+dt\right) = \hat{\textbf{U}}\textbf{L}\left(t_0\right)$ where $\hat{\textbf{U}} = e^{\textbf{J}*dt}$. Using the modified Gram-Schmidt \textbf{QR} decomposition on $\textbf{L}\left(t_0+dt\right)$, the diagonal elements of $\textbf{R}$ account for the Lyapunov exponents $\tilde{\lambda}_i\left(i=1,\dots, n\right)$ at time $t_0+dt$, that is
\begin{eqnarray}
\tilde{\lambda}_i(t_0+dt) = \frac{1}{dt}\ln\left(\textbf{R}_{ii}(t_0+dt)\right).
\end{eqnarray}
Repeating this procedure several time, after a large number of iterations $N$, the Lyapunov exponents can be approximated by
\begin{eqnarray}
\lambda_{i}\equiv\langle\tilde{\lambda}_i\rangle = \frac{1}{Ndt}\sum_{k = 1}^{N}\ln\left(\textbf{R}_{ii}(t_0+kdt)\right).
\end{eqnarray}

\section{Spatiotemporal chaos dimension $\xi_\delta$}
\label{app:lyap_dens_dim}
 Considering the extensive feature of the spatiotemporal chaos, the Kaplan-Yorke dimension,
\begin{equation}
D_{KY}=p +\frac{\sum_{i=1}^p \lambda_i}{\lambda_{p+1}},
\end{equation}
where $p$ is the largest integer that satisfies $\sum_{i=1}^p \lambda_i>0$. $D_{KY}$
may changes linearly with the volume of the system \cite{Ruelle1982,Cross1993}. That is, for a 1D system, $D_{KY}=\xi_\delta^{-1} \Delta T$ 
where $\Delta T$ is the extension of the system and $\xi_\delta$ represents the { Lyapunov dimension density} of the system for a {fixed} value of the control parameter. This quantity gives an estimation of the extension of the dynamically independent 
subsystems.

\section{Equal time correlation length $\xi_2$}
\label{app:corr2_dim}
The correlation length $\xi_2$  is defined as the exponential decay of the equal time two-point correlation \cite{Ohern1996,Egolf1994,Cross1993}:
\begin{equation}
\label{eq:2ptcorr}
C\left(\Delta\tau\right)=\langle\left(\psi\left(\Delta\tau+\tau^\prime,t\right)-\langle\psi\rangle\right)\left(\psi\left(\tau^\prime,t\right)-\langle\psi\rangle\right)\rangle,
\end{equation}
where the brackets  $\langle\cdot\rangle$ stand for the average process. The direct determination of $C\left(\Delta\tau\right)$ is quite costly in calculation time. However, by using the Wiener-Khintchin theorem \cite{Reif1975,Egolf1995}, it is computed by the following process: first time-averaging the Fourier spectra and next taking the inverse Fourier transform {of} its magnitude squared. Since the experimental spectra result from an averaging process over a large number of cavity round trip, $C\left(\Delta\tau\right)$ can also be computed taking the inverse Fourier transform of the measured spectrum.  Hence, for the LL equation {(\ref{eq:LL})}, we have computed $\xi_\delta$ end $\xi_2$ with respect to the input pump intensity.


\bibliography{biblio_sti}

\end{document}